\documentclass[suppldata]{interact}

\usepackage{epstopdf}% To incorporate .eps illustrations using PDFLaTeX, etc.
\usepackage[caption=false]{subfig}% Support for small, `sub' figures and tables
\usepackage[numbers,sort&compress]{natbib}

\usepackage{multirow}

\makeatletter% @ becomes a letter
\def\NAT@def@citea{\def\@citea{\NAT@separator}}% Suppress spaces between citations using natbib.sty
\makeatother% @ becomes a symbol again

\theoremstyle{plain}% Theorem-like structures provided by amsthm.sty

\usepackage{algorithm,algorithmic}

\theoremstyle{definition}

\theoremstyle{remark}

\date{}
\begin{document}

%\begin{titlepage}
%\title{Author detials}
%\maketitle
%full names, affiliations, postal addresses, telephone numbers and email addresses
%Zhengwei Wang \\
%Affiliation: Dublin City University\\
%Postal address: Insight Centre for Data Analytics, School of Computing, Dublin 9, Ireland\\
%E-mail: zhengwei.wang22@mail.dcu.ie\\
%
%Graham Healy\\
%Affiliation: Dublin City University\\
%Postal address: Insight Centre for Data Analytics, School of Computing, Dublin 9, Ireland\\
%E-mail: graham.healy@dcu.ie\\
%
%Alan F. Smeaton\\
%Affiliation: Dublin City University\\
%Postal address: Insight Centre for Data Analytics, School of Computing, Dublin 9, Ireland\\
%E-mail: alan.smeaton@dcu.ie\\
%
%Tomas E. Ward\\
%Affiliation: Dublin City University\\
%Postal address: Insight Centre for Data Analytics, School of Computing, Dublin 9, Ireland\\
%E-mail: tomas.ward@dcu.ie\\
%\thispagestyle{empty}
%\end{titlepage}

\articletype{Original Research}% Specify the article type or omit as appropriate

\title{Spatial Filtering Pipeline Evaluation of Cortically Coupled Computer Vision System for Rapid Serial Visual Presentation}

\author{
\name{Zhengwei Wang\thanks{This work has been accepted by Brain-Computer Interfaces. DOI:10.1080/2326263X.2019.1568821.}, Graham Healy, Alan F. Smeaton and Tomas E. Ward}
\affil{Insight Centre for Data Analytics, Dublin City University, Dublin 9, Ireland\\
$\texttt{Email: zhengwei.wang22@mail.dcu.ie}$, $\left\{\texttt{graham.healy, alan.smeaton, tomas.ward}\right\}$\texttt{@dcu.ie}}
}

\maketitle

\begin{abstract}
Rapid Serial Visual Presentation (RSVP) is a paradigm that supports the application of cortically coupled computer vision to rapid image search. In RSVP, images are presented to participants in a rapid serial sequence which can evoke Event-related Potentials (ERPs) detectable in their Electroencephalogram (EEG). The contemporary approach to this problem involves supervised spatial filtering techniques which are applied for the purposes of enhancing the discriminative information in the EEG data. In this paper we make two primary contributions to that field: 1) We propose a novel spatial filtering method which we call the Multiple Time Window LDA Beamformer (MTWLB) method; 2) we provide a comprehensive comparison of nine spatial filtering pipelines using three spatial filtering schemes namely, MTWLB, xDAWN, Common Spatial Pattern (CSP) and three linear classification methods Linear Discriminant Analysis (LDA), Bayesian Linear Regression (BLR) and Logistic Regression (LR). Three pipelines without spatial filtering are used as baseline comparison. The Area Under Curve (AUC) is used as an evaluation metric in this paper. The results reveal that MTWLB and xDAWN spatial filtering techniques enhance the classification performance of the pipeline but CSP does not. The results also support the conclusion that LR can be effective for RSVP based BCI if discriminative features are available.
\end{abstract}


\section{Introduction} \label{sec:introduction}
There is growing interest in using Electroencephalography (EEG) signals to help in searching images \cite{1642762, bigdely2008brain, healy2011eye}. This is based on estimating image content by examining participants' neural signals in response to image presentation. The concept of Rapid Serial Visual Presentation (RSVP) can be introduced using a familiar example, that of rapidly riffling through the pages of a book in order to locate a needed image \cite{spence2013rapid}. In RSVP, a rapid succession of target and standard (non-target) images are presented to a participant on a display at a rate of 5 $Hz$ - 12 $Hz$. The location of target images within the high-speed presentation is not known in advance by users and hence requires them to actively look out for targets i.e. to attend to target images. This paradigm where users are instructed to attend to target images amongst a larger proportion of standard images is known as an oddball paradigm and is commonly used to elicit Event-related Potentials (ERPs) such as the P300, a positive voltage deflection that typically occurs between 300 $ms$ - 600 $ms$ after the appearance of a rare visual target within a sequence of frequent irrelevant stimuli \cite{polich2007updating}. Since participants do not know when target images will appear in the presentation sequence, their occurrence causes an attentional-orientation response that is characterized by the presence of a P300 ERP. 
 %RSVP related EEG aims to develop a BCI application which is related to human beings' perception and visual system. 

Using modern signal processing and machine learning techniques, RSVP can be coupled with single-trial ERP detection to enable image search BCI applications \cite{lawhern2018eegnet, lees2018review}. Single-trial ERPs detection for an RSVP paradigm presents the following challenges:

%cb
\textit{Challenge 1. Low Signal-to-noise Ratio (SNR):}  ERP component amplitudes are often much smaller than those of spontaneous EEG components and task-related ERP components are typically overwhelmed by strong ongoing EEG background activity in single trials and so cannot be normally visually recognized in the raw EEG trace \cite{teplan2002fundamentals}. Traditional methods analyze ERPs by averaging across several task-related trials in order to reduce or eliminate spontaneous EEG components \cite{luck2012event}. 

%cb
\textit{Challenge 2. Curse of dimensionality:} RSVP-based ERP data can have high dimensionality spanning both space and time. Moreover, ERPs vary greatly across participants and experimental task \cite{luck2012event}. In order to capture the ERPs, it is necessary to choose a time window large enough for epoching which involves the time region in which ERPs might appear. Moreover the training sets available for machine learning purposes are typically modest in size and worse, contain relatively few instances of the responses evoked by the infrequent (by design) target image class.

%cb
\textit{Challenge 3. Overlapping epochs:} The strength of the RSVP paradigm is that the rate of the stimulus sequence increases the upper limit of potential information transfer rates in BCI applications. However, a relatively large time window has to be set for epoching in order to capture the ERPs. Therefore, there is substantial overlap between adjacent target epochs and standard epochs because of the short Interstimulus Interval (ISI) used in the RSVP paradigm.

In this work, we consider a pipeline combining spatial filtering and linear classification as this is the most widely used pipeline in RSVP-EEG based BCI.

There are several potential signal pre-processing techniques that may increase the efficiency of detecting single-trial ERPs including time-frequency feature extraction and hierarchical discriminant component analysis (HDCA) \cite{meng2012characterization, parra2008spatiotemporal}. However, spatial filtering techniques are more efficient when a full EEG cap dataset is available. In this paper, we focus on spatial filtering for signal pre-processing as this is the predominant approach used in RSVP-based BCI because we are using a full EEG cap recording dataset. Spatial filtering focuses on enhancing task-related information contained in the EEG signal. It plays an important role in BCI research because it can enhance the discriminative information present in the EEG signal whilst reducing the overall data dimensionality. Spatial filtering has been shown to enhance detection accuracy with a P300 speller paradigm \cite{rivet2009xdawn}. However, classification pipelines without spatial filtering have been proposed for  single-trial ERP detection. These methods include widely used linear classifiers such as Linear Discriminant Analysis (LDA) \cite{meng2012characterization}, Logistic Regression (LR) \cite{huang2006boosting} and Bayesian Linear Regression (BLR) \cite{hoffmann2008efficient}.

Investigation of spatial filtering in RSVP-based EEG has been explored previously in the literature \cite{cecotti2014single, cecotti2017best, wang2018review}. What is unclear from these studies is how to determine the optimal number of spatial filters i.e. this detail has been omitted in previous studies and yet this is an important consideration so is included in this investigation. The primary objective of this paper is to explore the performance of pipelines that combine different spatial filtering approaches and classifiers where their respective hyperparameters are explored through random search cross validation. A pipeline in this paper comprises spatial filtering, feature dimensionality reduction and classification steps. Three spatial filtering approaches are explored in this paper, namely xDAWN \cite{rivet2009xdawn}, multiple time window LDA beamformers (MTWLB) which is an extension of LDA beamformer \cite{treder2016lda} and common spatial pattern (CSP) \cite{blankertz2008optimizing}. Principal component analysis (PCA) is utilized for feature dimensionality reduction. Three linear classification methods are explored, namely LDA, BLR and LR respectively. There are nine pipelines in total including spatial filtering and classification combinations. Three pipelines that only apply the three classification methods without applying any spatial filtering are used for comparing performance. This paper should provide neurotechnologists seeking to apply a RSVP paradigm to EEG with a comprehensive assessment of the comparative performance of both commonly used spatial filtering pipelines and a new method all assessed using a new publicly available benchmark dataset.

This paper is organized as follows. Firstly, we describe the methodology which includes pipeline and performance evaluation metrics used in this paper. Secondly, we clarify the experimental RSVP-based EEG dataset used in this paper. Finally, results and discussion are presented in the last two sections.

\section{Methodology}\label{class}
\subsection{Pipeline Description}
\begin{figure*}[h!]
    \centering
    \subfloat[Pipeline including spatial filtering]{
    	\includegraphics[scale=0.35]{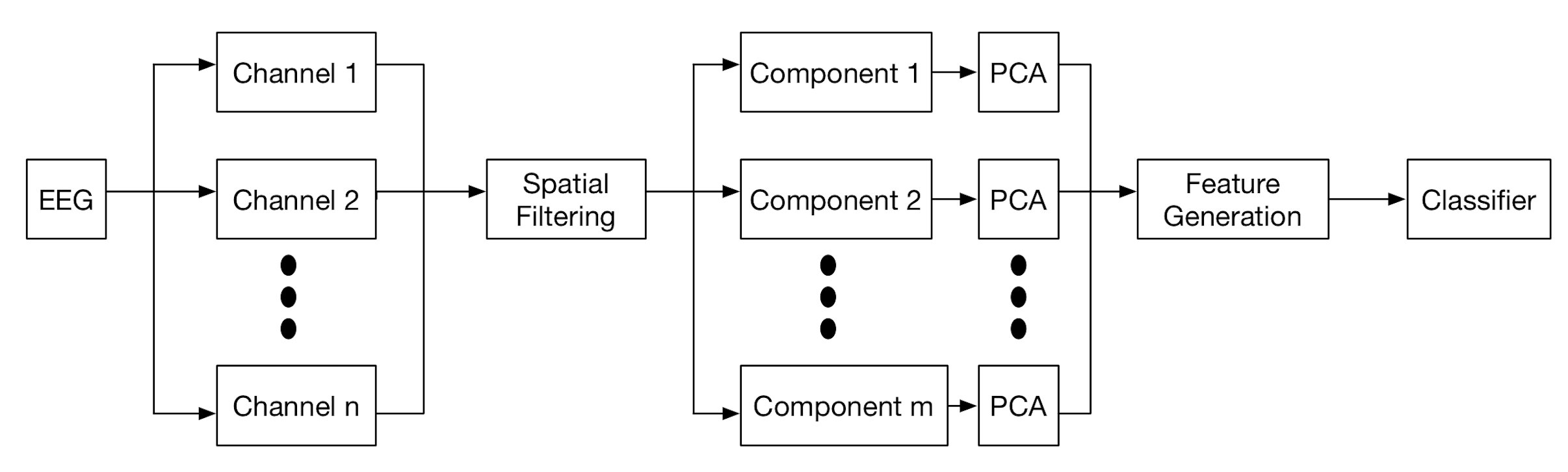}
     	\label{fig:pipeline_SF}
    }
    \hfill
    \centering
    \subfloat[Pipeline excluding spatial filtering]{
    	\includegraphics[scale=0.3]{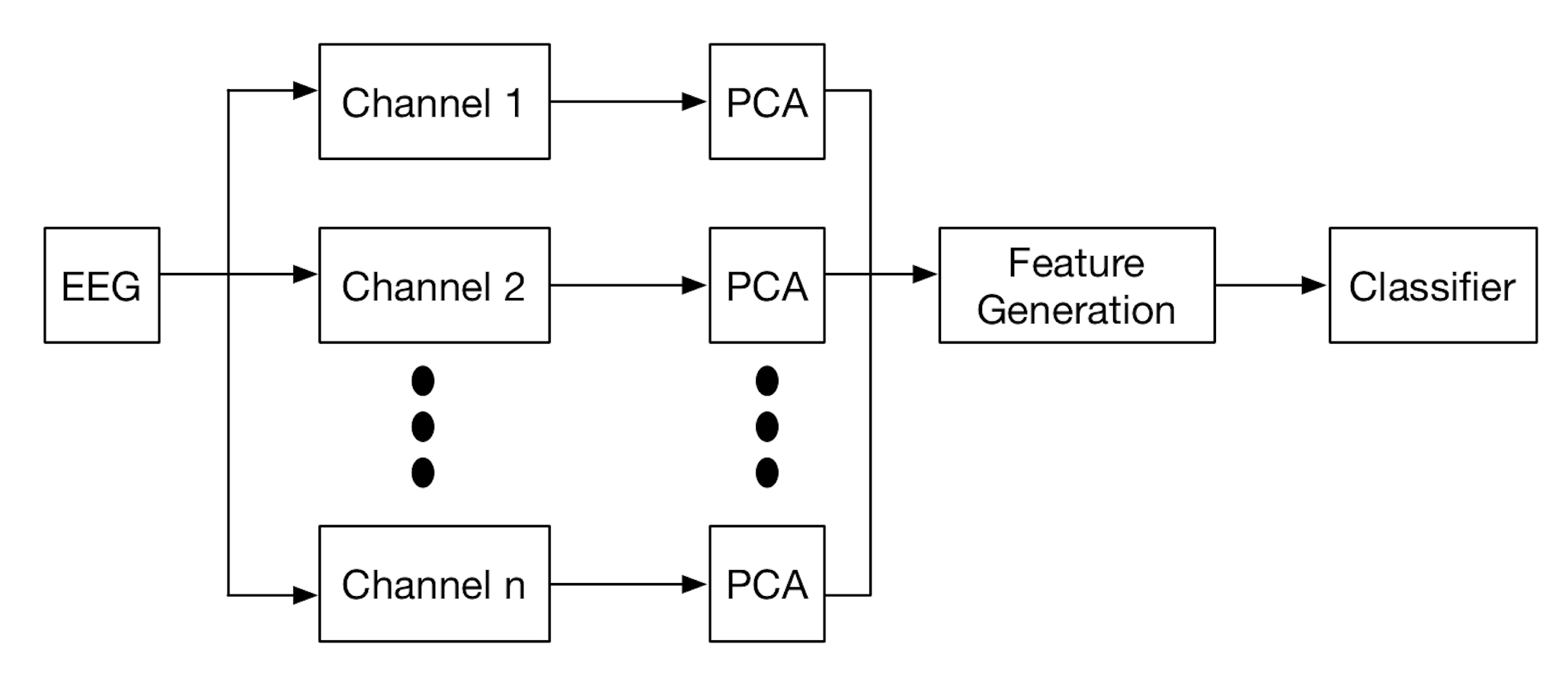}
     	\label{fig:pipeline}
    }
    
   \caption{Two pipeline architectures for RSVP-based EEG discussed in this paper. }
   \label{fig:two-pipelines}
\end{figure*}

This paper explores nine pipelines comprising spatial filtering, feature dimensionality reduction and classification respectively along with three pipelines containing feature dimensionality reduction and classification as comparisons. Fig.~\ref{fig:two-pipelines} illustrates the two pipeline architectures under consideration in this study. With spatial filtering applied, an $n$ channel EEG epoch is transformed to $m$ source components ($m < n$). Before the feature generation step, we apply PCA for each individual channel $\in \mathbb{R}^{L \times n}$ (pipelines without spatial filtering) and individual component $\in \mathbb{R}^{L \times m}$ (pipelines with spatial filtering) on the temporal axis for dimensionality reduction following the work \cite{alpert2014spatiotemporal}, where $L$ is the number of epochs. The reason why we apply PCA individually is because EEG power in each channel and each component is not consistent and this step ensures that discriminative information is not lost. We leave out the PCA components which contain less than 1\% ratio of the variance. The feature generation step concatenates $m$ components or $n$ channel EEG to a feature vector before feeding to the classifier.

\subsection{Supervised Spatial Filtering}
Before introducing the supervised spatial filtering approaches we clarify the notations which will be used. $N_{c}$ is the number of channels, $N_{t}$ is the number of time samples in an epoch, $N_{f}$ is the number of selected spatial filters and $n$ is the number of epochs.

Spatial filtering creates a weighted combination of each EEG channel input in order to enhance a particular subset of information which is contained in the original EEG epoch. Spatial filtering reduces the number of features because the number of spatial filters selected $N_{f}$ is smaller than the number of channels $N_{c}$.
% changed for reviewer 1 major concern 4: delete the problem formulation
The problem of spatial filtering is to find projection vectors (spatial weights for each channel) $\mathbf{w}\in \mathbb{R}^{N_{c}\times N_{f}}$ to project $\mathbf{X} \in \mathbb{R}^{N_{c}\times N_{t}}$ to a subspace, where $\mathbf{w}$ is calculated by different optimization criterion.
\begin{equation} \label{eq:spatial-filtering-formulation}
	\mathbf{X}_{\text{sub}}=\mathbf{w}^{T}\mathbf{X}
\end{equation}

Several approaches have been presented in the literature for generating spatial filters $\mathbf{w}$ in equation \eqref{eq:spatial-filtering-formulation} in the area of BCI research. Independent component analysis (ICA) is a blind source separation technique which can be used to find a linear representation of non-Gaussian data so that the components are statistically independent, or as independent as possible \cite{ hyvarinen2000independent}. Such a representation is capable of capturing the inherent structure of data in many applications, and hence has application to feature extraction \cite{makeig2009erp, bigdely2008brain} and removing artifacts from EEG signals \cite{jung2000removing}. Specifically, ICA finds a component `unmixing' matrix ($\mathbf{w}$) that, when multiplied by the original data ($\mathbf{X}$), yields the matrix ($\mathbf{X}_{sub}$) of independent component (IC) time courses \cite{onton2006imaging}. Therefore, the main purpose of ICA is blind source separation instead of discriminating EEG in two experimental tasks. PCA is another statistical technique that uses eigenvalue decomposition to convert a set of correlated variables into a set of linearly uncorrelated variables. PCA has been applied to EEG signals for dimensionality reduction \cite{naeem2009dimensionality} and generating spatial filters \cite{zanotelli2010optimum}. Similar to ICA, PCA operates without knowledge of stimulus types hence it is an unsupervised approach. In this work, we consider supervised spatial filtering methods that aim to enhance the difference between target and standard image stimuli. Three spatial filtering methods are considered in this paper, namely MTWLB, xDAWN and CSP. MTWLB is an extension of the LDA beamformer method which aims to maximize the signal-to-noise ratio (SNR) in each individual time window. xDAWN and CSP are based on Rayleigh quotients where xDAWN maximizes the signal to signal plus noise ratio (SSNR) whereas CSP maximizes the difference of the variance between two classes. In the following paragraphs, we describe the operation of these three spatial filters generation techniques, i.e. The LDA beamformer with our window extensions, xDAWN and CSP.

\subsubsection{LDA Beamformer}
The LDA beamformer has been successfully applied for recovering N2 and P3 sources in an auditory experiment \cite{treder2016lda}. Considering the epoch $\mathbf{X} \in \mathbb{R}^{N_{c} \times N_{t}}$, let column vectors $\mathbf{p_{1}}$ $\in$ $\mathbb{R}^{N_{c}\times 1}$ and $\mathbf{p_{2}}$ $\in$ $\mathbb{R}^{N_{c}\times 1}$ be the spatial pattern of a specific component in two different experimental conditions. We denote the difference pattern as $\mathbf{p}:=\mathbf{p_{1}}-\mathbf{p_{2}}$ and the covariance matrix $\mathbf{\Sigma}$ $\in$ $\mathbb{R}^{N_{c}\times N_{c}}$. The optimization problem of the LDA beamformer can be stated as
% changed for the reviewer 1 minor concern 3
\begin{equation}\label{eq:lda-beamformer}
\begin{gathered}
  \mathop{\text{minimize}}_{\mathbf{w}}\hspace{5pt} {\mathbf{{w}}^{\text{T}} \mathbf{\Sigma} \mathbf{w} } \hspace{10pt}
  \text{s.t.} \hspace{2pt} \mathbf{w}^{\text{T}} \mathbf{p} =1
\end{gathered}
\end{equation}
where $\mathbf{w}$ is the spatial filter. Consequently, the solution to the optimization problem maximizes the signal-to-noise ratio (SNR) of the desired signal and the optimal spatial filter is given by
\begin{equation}\label{eq:LDA-beamformer-solution}
  \mathbf{w}=\mathbf{\Sigma}^{-1} \mathbf{p} ( \mathbf{p}^{\text{T}} \mathbf{\Sigma}^{-1} \mathbf{p}) ^{-1}
\end{equation}

\begin{figure}[h!]
    \centering
    \includegraphics[scale=0.6]{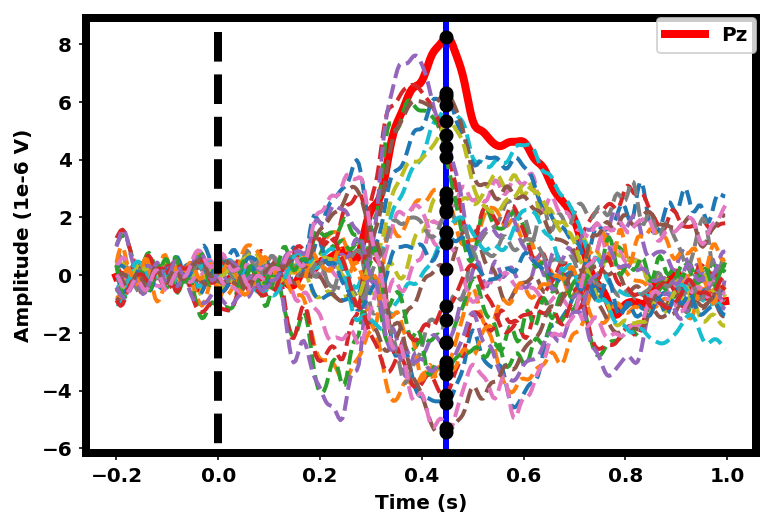}
    \caption{Spatial pattern estimation for LDA beamformer using whole EEG epoch via training data using CAR: \textit{Participant 2}.} \label{fig:pattern-estimation-beamformer}
\end{figure}

The spatial pattern for the LDA beamformer is directly estimated from the difference between ERP peaks in an oddball experiment \cite{treder2016lda}. As seen in Fig.~\ref{fig:pattern-estimation-beamformer}, the bold red line is the ERP difference at the Pz channel and the blue line represents the peak value timestamp of difference ERPs at the Pz channel. The different ERP values across all channels at that timestamp can then be estimated as spatial patterns.

Due to the substantial overlap between adjacent target epochs and non-target epochs, along with inherent variability in ERP latencies and topographies between participants, we extend the LDA beamformer to MTWLB. The MTWLB implementation can be seen in Algorithm \ref{alg:MTWLB}. The key idea of MTWLB is to train multiple LDA beamformer models over non-overlapping successive time windows i.e. to train a single LDA beamformer for each time window that is adaptive to the local spatio-temporal features characterizing target-related ERP activity at that time point.

\begin{algorithm}
\begin{algorithmic}[1]
\caption{Implementation of MTWLB} \label{alg:MTWLB}
	\scriptsize
    \STATE Set $M$ time windows for MTWLB;
    \STATE Calculate $\mathbf{\Sigma}$ in each time window:
    \FOR {$i=1:M$}
    \FOR{Each difference ERP $\mathbf{p}$ at each timestamp in an individual time window}
        \STATE $\mathbf{w}=\mathbf{\Sigma}^{-1} \mathbf{p} ( \mathbf{p}^{\text{T}} \mathbf{\Sigma}^{-1} \mathbf{p}) ^{-1}$
        \STATE $J=\mathbf{w}^{\text{T}} \mathbf{\Sigma} \mathbf{w}$
    \ENDFOR
    \STATE Retain $\mathbf{w}$ and $\mathbf{p}$ that minimize $J$
    \ENDFOR

\end{algorithmic}
\end{algorithm}
 
\paragraph{Spatial pattern estimation for MTWLB}
In contrast with the LDA beamformer method, we estimate the spatial pattern and calculate the equation \eqref{eq:LDA-beamformer-solution} separately for each time window rather than whole time series. Therefore, there are $M$ sets of estimated spatial patterns while each set has 32 individual spatial patterns. The reason why we do this is the substantial overlap between adjacent target epochs and non-target epochs, along with inherent variability in ERP latencies and topographies between participants in RSVP-based EEG.

\paragraph{Covariance matrix estimation for MTWLB}
MTWLB uses the specified time window coinciding with ERP activity to  
estimate the covariance matrix instead of using the whole EEG epoch that is utilized by LDA beamformer \cite{treder2016lda}. We firstly concatenate $n$ trial EEG epochs through time and trials (i.e. the concatenated data $\mathbf{D} \in \mathbb{R}^{N_{c}\times N}$, where $N=N_{t}\times n$) and then normalize the concatenated data through columns. The covariance matrix $\mathbf{C} \in \mathbb{R}^{N_{c}\times N_{c}}$ can be calculated from the normalized concatenated data. After using artifact rejection or other EEG pre-processing  methods, the covariance matrix is singular. We used shrinkage algorithms to regularize the covariance matrix in order to make it invertible \cite{chen2010shrinkage}.

As a result, MTWLB is able to generate a set of spatial filters corresponding to ERP differences in each specified time window where the number of time windows can be selected using cross-validation.

\subsubsection{xDAWN}
The xDAWN algorithm has been applied  in BCI for ERP detection in the P300 speller paradigm \cite{cecotti2011robust, rivet2009xdawn} and the RSVP paradigm \cite{cecotti2014single}. It considers the model of the recorded signals $\mathbf{X} \in \mathbb{R}^ {N_{t} \times N_{c}}$ as follows:
\begin{equation} \label{eq:xDAWN-model}
	\mathbf{X}=\mathbf{D} \mathbf{A} + \mathbf{N}
\end{equation} 
where $\mathbf{D} \in \mathbb{R}^{N_{t} \times N_{e}}$ ($N_{e}$ is the number of temporal samples of the ERP response) is the Toeplitz matrix whose first column elements are set to zero except for those corresponding to a target onset, which are set to one, $\mathbf{A} \in \mathbb{R}^{N_{e} \times N_{c}}$ is the matrix of ERPs. Hence, $\mathbf{D} \mathbf{A}$ in \eqref{eq:xDAWN-model} represents the ERP response corresponding to the synchronous response with target stimuli. $\mathbf{N}$ is the on-going brain activity, also known as EEG background noise. 

The goal of xDAWN is to apply spatial filters $\mathbf{w}$ to enhance the SSNR of the ERP response corresponding to the target stimulus. The optimization problem for xDAWN can be defined as
\begin{equation} \label{eq:xDAWN-solution}
	\mathbf{\hat{w}}=\arg\max_{\mathbf{w}} \frac{\text{Trace}(\mathbf{w}^{T} \mathbf{\hat{A}}^{T} \mathbf{D}^{T} \mathbf{D} \mathbf{\hat{A}} \mathbf{w})}{\text{Trace}(\mathbf{w}^{T} \mathbf{X}^{T} \mathbf{X} \mathbf{w})}
\end{equation}
where $\mathbf{\hat{A}}$ is the least squares estimation of $\mathbf{A}$ \cite{rivet2009xdawn}. Thus, a number of spatial filters corresponding to different evoked responses can be obtained through the Rayleigh quotient optimzation problem \cite{parlett1998symmetric}. The number of spatial filters are often chosen through cross-validation.

\subsubsection{Common Spatial Pattern}
CSP is one of the most popular spatial filtering approaches for motor imagery based BCIs, where the task involves two different states of brain activity (e.g. imagery of the movement of the left or right hand) \cite{blankertz2008optimizing, wang2006common}. CSP aims to maximize the variance of one class and minimize the variance of another class. The optimization problem for CSP can also be estimated and interpreted as Rayleigh quotient \cite{parlett1998symmetric}.

First, let $\mathbf{X}_{1}(i)$ and $\mathbf{X}_{0}(i)$ be the $i$th event locked ERP epoch $\in \mathbb{R}^{N_{c} \times N_{t}}$ and two covariance matrices $\mathbf{\Sigma_{1}}$ and $\mathbf{\Sigma_{0}}$ are calculated as follows (subscript ``0'' for standard condition and ``1'' for target condition)  
\begin{equation}
	\mathbf{\Sigma_{a}}= \frac{1}{n} \sum_{i=1}^{n}  \frac{\mathbf{X}_{a}(i) \mathbf{X}_{a}^{T}(i)}{\text{Trace}(\mathbf{X}_{a}(i) \mathbf{X}_{a}^{T}(i))}
\end{equation}

The solution for CSP can be determined through Raleigh quotients by solving a generalized eigenvalue problem
\begin{equation} \label{eq:CSP-solution}
	\left\{ \text{max, min}\right\}_{\mathbf{w}} \hspace{6pt} \frac{\mathbf{w}^{T} \mathbf{\Sigma_{1}} \mathbf{w}}{\mathbf{w}^{T} \mathbf{\Sigma_{0}} \mathbf{w}} 
\end{equation}

% changed for reviewer 1 minor concern 4
Similar to the previous two approaches, CSP is able to generate a set of spatial filters. However, spatial filters in CSP appear pair-by-pair because CSP maximizes variance in one class and minimize variance in the other class. From Cecotti's work, four spatial filters were chosen as $\left[\mathbf{w}_{1}, \mathbf{w}_{2}, \mathbf{w}_{N_{c}-1}, \mathbf{w}_{N_{c}} \right]$ ($N_{c}$ is the number of electrodes) \cite{cecotti2014single}. This work chooses a pair of spatial filters via cross-validation. 

% changed for reviewer 1 minor concern 5
At this point, we have highlighted how the three methods under consideration here can generate spatial filters $\mathbf{w}$. All three spatial filters generated serve the same objective of reducing computation complexity but the optimization target are different. MTWLB generates spatial filters based on maximizing the SNR in individual time windows. xDAWN, in contrast generates spatial filters based on maximizing the signal-to-signal plus noise ratio (SSNR) for the whole EEG epoch. Finally the method of CSP generates spatial filters through maximizing the variance difference between two classes.

\subsection{Feature Generation}
The spatial filter $\mathbf{w} \in \mathbb{R}^{N_{c} \times N_{f}}$ generated serves to transform the original EEG epoch $\mathbf{X} \in \mathbb{R}^{N_{c} \times N_{t}}$ to the feature space
\begin{equation} \label{Transform}
	\mathbf{\Psi}=\mathbf{w}^{\text{T}} \mathbf{X}
\end{equation}
where $\mathbf{\Psi} \in \mathbb{R}^{N_{f} \times N_{t}}$. The projected subspace $\mathbf{\Psi}$ can be represented as spatial-filtered EEG signals involving different discriminative information corresponding to the criteria used in their filter generation. PCA can then be applied to each row in $\Psi$ for feature reduction and generating a new set of time series which are linearly uncorrelated. In this work, principal components whose explained variance ratio is greater than 1\% are selected and concatenated as the feature vector which will be used as input to the classification step.

\subsection{Linear Classifiers}
Linear classifiers are widely used for RSVP-based BCI due to their good performance, often simple implementation and low computational complexity \cite{bigdely2008brain, meng2012characterization, 1642762, parra2008spatiotemporal}. In this paper, we  focus on three widely used linear classifiers in RSVP-based BCI research, namely LDA, LLR and BLR.

\subsubsection{Linear Discriminant Analysis}    
LDA is a supervised subspace learning method which is based on the Fisher criterion and it is equivalent to least squares regression (LSR) if the regression targets are set to $\frac{N}{N_{1}}$ for samples from class 1 and $\frac{N}{N_{2}}$ for samples from class 2 (where $N$ is total number of training samples, $N_{1}$ is the number of samples from class 1 and $N_{2}$ is the number of samples from class 2) \cite{bishop2006pattern}. It aims to find an optimal linear transformation $\mathbf{w}$ that maps $\mathbf{x}$ to a subspace in which the between-class scatter is maximized while the within-class scatter is minimized in that subspace. The optimization problem for LDA is to maximize the cost function as below
\begin{equation}
	J={ {\mathbf{w}^{T}\mathbf{S}_{B}\mathbf{w} } \over { \mathbf{w}^{T}\mathbf{S}_{W}\mathbf{w} } }
\end{equation}
where $J$ represents the cost to be minimized, $\mathbf{S}_{B}$ is the between-class scatter, $\mathbf{S}_{W}$ is the within-class scatter, $\mathbf{w}^{T}$ is the transpose matrix to the $\mathbf{w}$. Regularization is often applied in order to avoid the singular matrix problem of $\mathbf{S}_{W}$ \cite{friedman1989regularized}. LDA enables the best separation between two classes on the subspace $\mathbf{w}$. LDA has relatively low computational complexity which makes it suitable for online BCI systems. As mentioned earlier, classification of RSVP-based EEG data suffers from the imbalanced data set problem. In Xue's work \cite{xue2008unbalanced}, he showed that there is no reliable empirical evidence to support that an imbalanced data set has a negative effect on the performance of LDA for generating the linear transformation vector. Consequently, LDA is suitable and has been successfully used in RSVP-based BCI \cite{bigdely2008brain, meng2012characterization}.

\subsubsection{Bayesian Linear Regression}
Bayesian linear discriminant analysis (BLDA), can be seen as an extension of LDA or LSR. In BLR, regularization for parameters is used to prevent overfitting caused by high dimensional and noisy data. BLR assumes the parameter distribution and target distribution are both Gaussian \cite{bishop2006pattern}. We  introduce LSR as a starting point for the description of BLR. The solution for LSR can be stated as
\begin{equation}
	\mathbf{w}=(\mathbf{X}\mathbf{X}^{\text{T}})^{-1}\mathbf{X}\mathbf{y}
\end{equation} 
Note that $y=\frac{N}{N_{1}}$ for class 1 and $y=-\frac{N}{N_{2}}$ for class 2 here (threshold can be determined by adding a column with all one as the first column in $\mathbf{X}$). LSR does not consider the parameter distribution in this case and it maximizes the likelihood. For BLR, it considers the parameter distribution and maximizes the posterior. Given the prior target distribution $p\mathbf(y) \sim \mathcal{N}(\mu,\beta^{-1})$ and parameter distribution $p(\mathbf{w}) \sim \mathcal{N}(0,\alpha^{-1}\textbf{I})$ (where $\beta$ and $\alpha$ are the inverse variance), BLR gives the optimized estimation for the parameter
\begin{equation}
	\mathbf{w}=\beta(\beta \mathbf{X}\mathbf{X}^{\text{T}}+\alpha\textbf{I})^{-1}\mathbf{X}\mathbf{y}
\end{equation} 
The optimization problem of BLR can be concluded as the maximum a posterior (MAP) estimation \cite{gauvain1994maximum}, which lies in the assumption of an appropriate prior distribution of the parameter to be estimated. Hence, the optimization depends on the hyperparameters $\beta$ and $\alpha$. In real-world applications, the hyperparameters can be tuned  using cross validation or the maximum likelihood solution with an iterative algorithm \cite{bishop2006pattern, mackay1992bayesian}. BLR has been shown to outperform LDA in BCI research \cite{hoffmann2008efficient, cecotti2017best}.

\subsubsection{Logistic Regression\label{Logistic regression}}
LR models the conditional probability as a linear regression of feature inputs. The logistic model can be constructed as
\begin{equation}
	p(\mathbf{x})={1\over{1+e^{-\mathbf{w}^{T}\mathbf{x}+b}}} \label{eq:LR_model}
\end{equation}
The optimization problem of an LR can be constructed by minimizing the cost function as below:
\begin{equation}  \label{eq:LR_cost}
\begin{split}
    J(\mathbf{w}, b)=
    &-{1\over{m}} \sum\limits_{i=1}^{m} { [y_{i}log(p(\mathbf{x}_{i}))+(1-y_{i})log(1-p(\mathbf{x}_{i})) ]}\\
    & + \lambda \mathbf{w}^{T} \mathbf{w}
\end{split}
\end{equation}
where $y\in \left\{0,1\right\}$, $m$ is the sample number of two classes and $\lambda$ is the regularization parameter  ($\lambda > 0$). 

LR is part of a broader family of generalized linear models (GLMs), where the conditional distribution of the response falls in some parametric family, and the parameters are set by the linear predictor. LR is the case where the response (i.e. $y$ in equation \eqref{eq:LR_cost}) is binomial and it can give the prediction of the conditional probability estimation. LR is easily implemented and has been successfully applied to RSVP based BCI research \cite{sajda2007single, huang2006boosting}.

% changed for the reviewer 1 minor concern 1
\subsection{Evaluation}
The evaluation described in this work seeks to assess the relative performance when combining three spatial filtering approaches with three linear classification methods, thus there are nine pipelines (spatial filtering $\times$ feature generation $\times$ classification) in total that are discussed in this paper. For comparison, the original EEG epochs without spatial filtering and only using PCA, are fed to three linear classifiers. Performance of the different pipelines are evaluated through the area under the curve (AUC) based on true positive rate (TPR) and false positive rate (FPR)

It should be noted that the pipeline described in this paper contains a number of hyperparameters. Three spatial filtering approaches contain a number of spatial filters $N_{f}$ as the hyperparameter. BLR contains data distribution variance ($\beta$) and parameter distribution variance ($\alpha$) while LR has the regularization term ($\lambda$) as hyperparameters. Only LDA does not require a hyperparameter. Table~\ref{table:hyperparameter-summary} summarizes the hyperparameters used in each pipeline. We apply a random search \cite{bergstra2012random} for 100 hyperparameter combinations in each pipeline and select these for evaluation on a test set using 10-fold cross validation. The optimal model is then applied to the testing data to calculate the AUC score.

\section{Data acquisition and pre-processing}
\begin{figure}[!htp]
	
	\subfloat{
		\includegraphics[width=.2\linewidth]{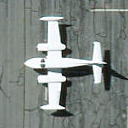}
	}
	\hfill
	\subfloat{
	
		\includegraphics[width=.2\linewidth]{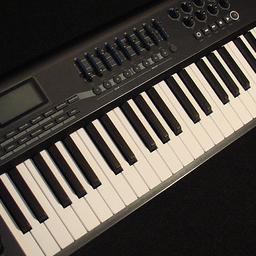}
		}
	\hfill
	\subfloat{
		\includegraphics[width=.2\linewidth]{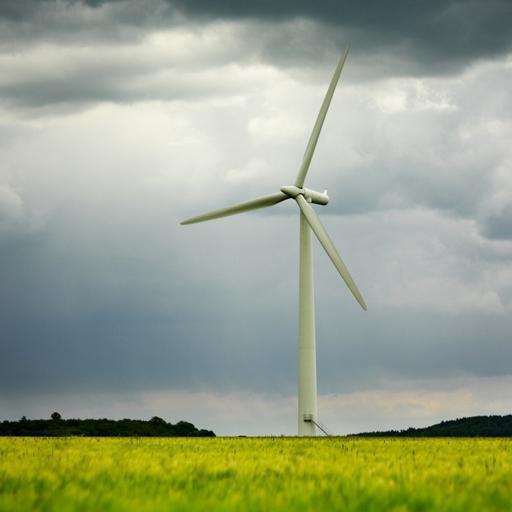}
	}
	\hfill
	\subfloat{
		\includegraphics[width=.2\linewidth]{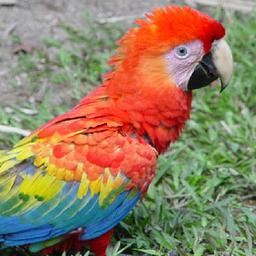}
	}
	\caption{Examples of four target images used in the experiment. Airplane, Keyboard instrument, Wind farm and Macaw respectively from left to right.}
	\label{fig:target-image}
\end{figure}

The EEG datasets used in this work is from the Neurally Augmented Image Labelling Strategies (NAILS) task as part of an open data challenge carried out in 2017 \cite{Graham}. EEG data from up to 9 participants in NAILS was used in this work. Data collection was carried out with approval from Dublin City University's Research Ethics Committee (DCU REC/2016/099). EEG was recorded along with timestamping information for image presentation (via a photodiode and hardware trigger) to allow for precise epoching of the EEG signals for each trial  \citep{wang2016investigation}. Each participant completed 6 different tasks (INSTR, WIND1, WIND2, UAV1, UAV2 and BIRD). For each task, participants were asked to search for specific target images from the presented images (i.e. an airplane has the role of target in UAV1 and UAV2 tasks, a keyboard instrument is the target for the INSTR task, while a windfarm is the target in the WIND1 and WIND2 tasks, parrot being the target in the BIRD task, see Fig.~\ref{fig:target-image}). Each task was divided into 9 blocks, where each block contained 180 images (9 targets/171 standards) thus there were 486 target and 9,234 standard images available for each participant. Images were presented to participants at a 6 Hz presentation rate. EEG data was recorded using a 32 channel BrainVision actiCHamp at 1000 $Hz$ sampling frequency, using electrode locations as defined by the 10-20 system.

Pre-processing of some kind is generally a required step before any meaningful interpretation or use of any EEG data can be realized. Pre-processing typically involves re-referencing, filtering the signal (by applying a bandpass filter to remove environmental noise or to remove activity in non-relevant frequencies), epoching (extracting a time epoch typically surrounding the stimulus onset) and trial/channel rejection (to remove those containing artifacts)  \cite{luck2012event}. In this work, a common average reference (CAR) was utilized and a bandpass filter (e.g. 0.1-30 $Hz$) was applied to the dataset. EEG data was then resampled at 250 $Hz$ and the analysis of a behavior response is considered between 0 and 1 $s$ after the presentation of a stimulus. Trial rejection based on HEOG and VEOG channels was applied for all participants. The dataset was split into a training/testing set of 66\%/33\% respectively, by selecting 3 blocks from each search task to act as a withheld test set in the evaluation.

\section{Results}
\subsection{Impact of Number of Spatial Filters}
\begin{figure}[!htp]
 \centering
	\subfloat[Early time region ERP for 9 participants from left to right.]{
    \centering
     \includegraphics[width=.09\linewidth]{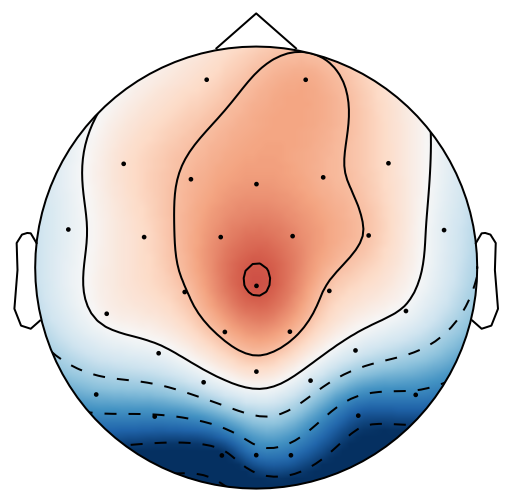}
     \includegraphics[width=.09\linewidth]{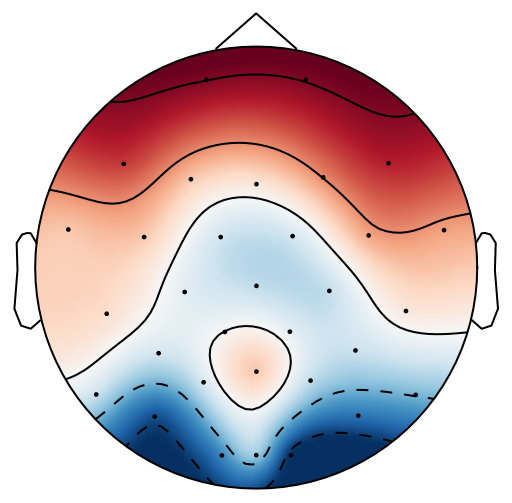}
     \includegraphics[width=.09\linewidth]{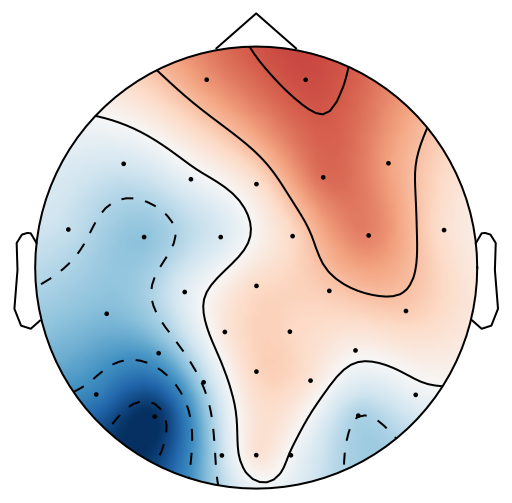}
     \includegraphics[width=.09\linewidth]{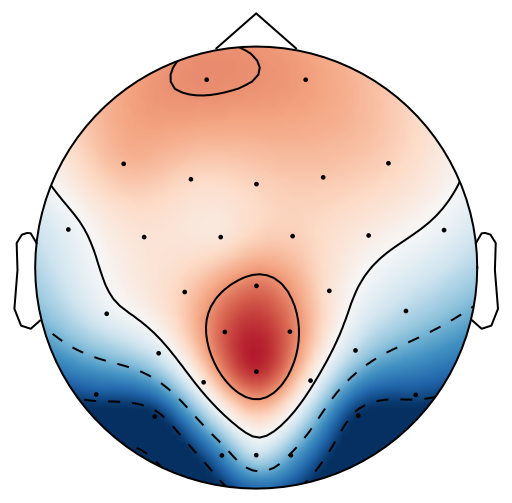}
     \includegraphics[width=.09\linewidth]{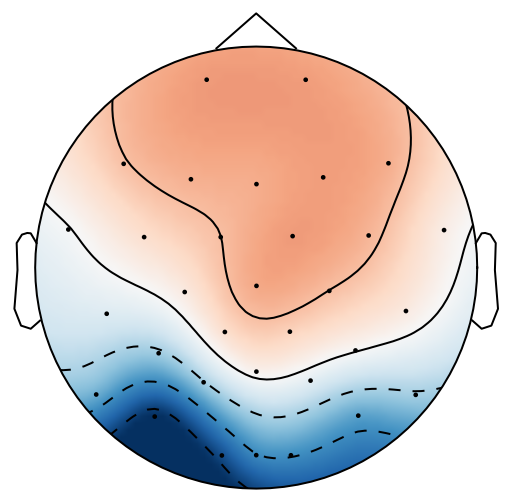}
     \includegraphics[width=.09\linewidth]{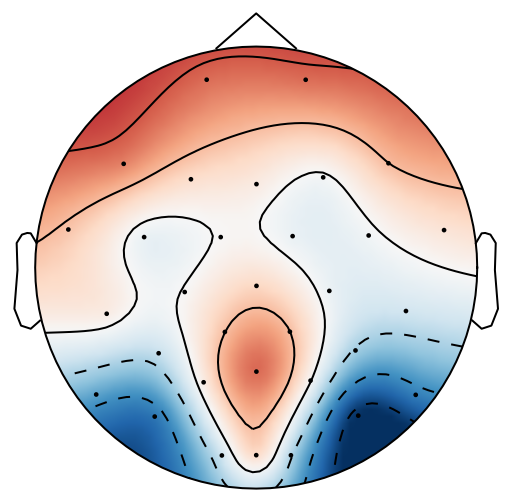}
     \includegraphics[width=.09\linewidth]{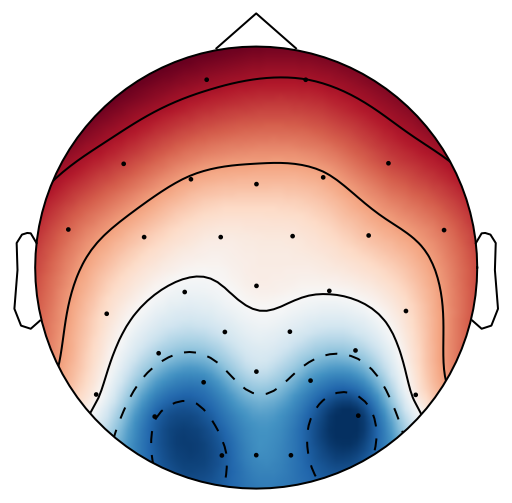}
     \includegraphics[width=.09\linewidth]{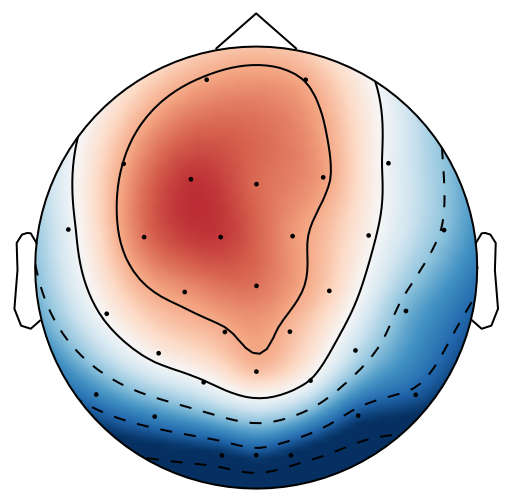}
     \includegraphics[width=.105\linewidth]{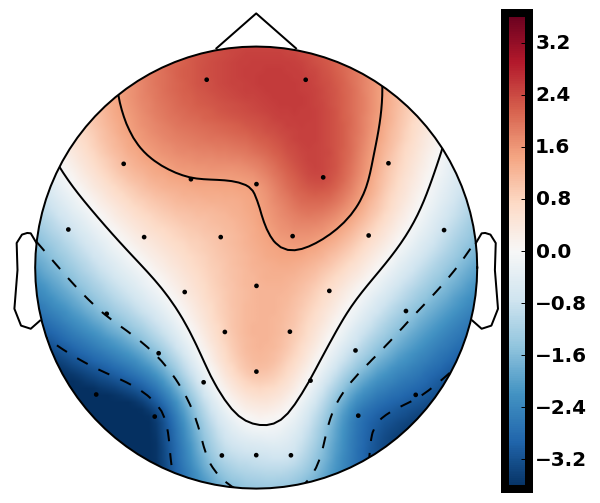}
     \label{fig:early-ERP}
	}

    \subfloat[Later time region ERP for 9 participants from left to right.]{
    \centering
     \includegraphics[width=.09\linewidth]{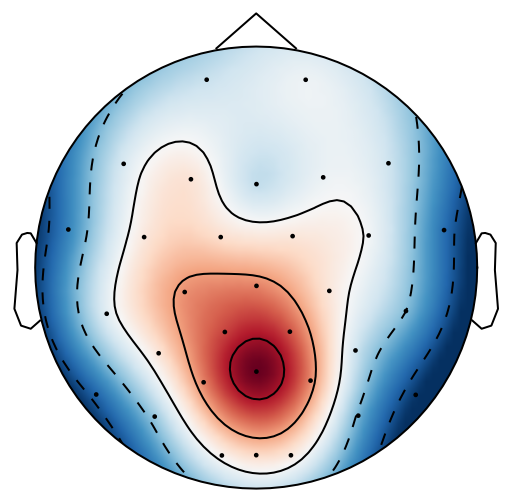}
     \includegraphics[width=.09\linewidth]{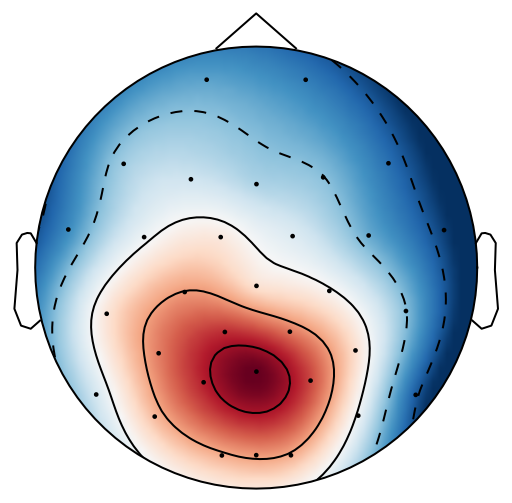}
     \includegraphics[width=.09\linewidth]{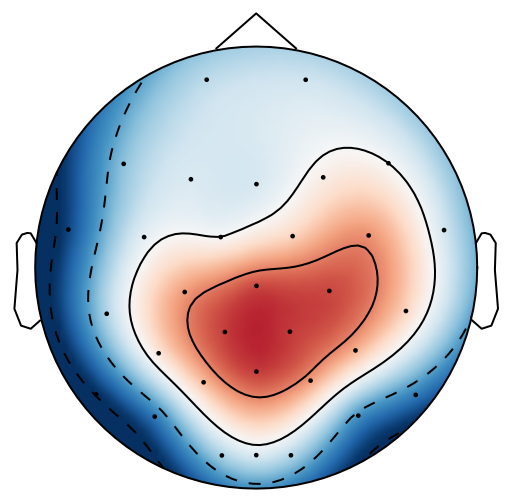}
     \includegraphics[width=.09\linewidth]{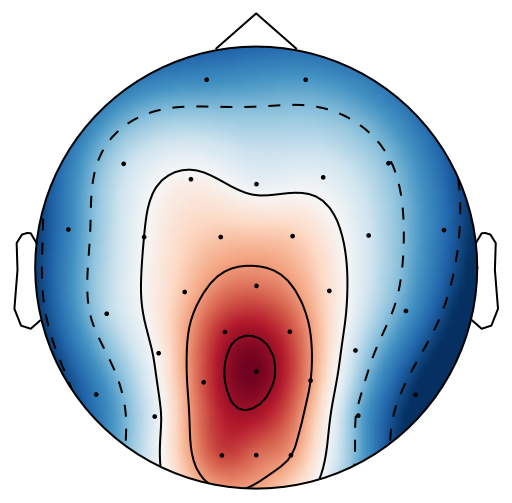}
     \includegraphics[width=.09\linewidth]{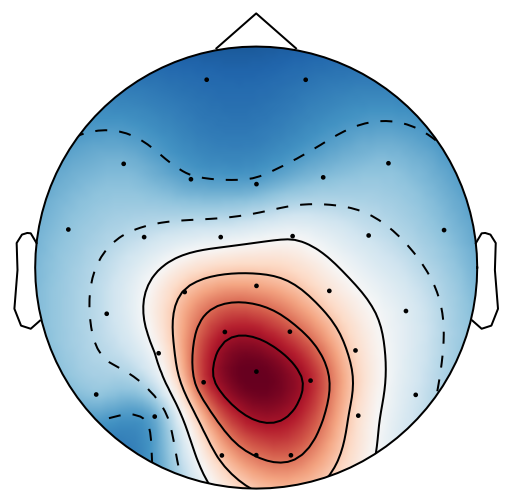}
     \includegraphics[width=.09\linewidth]{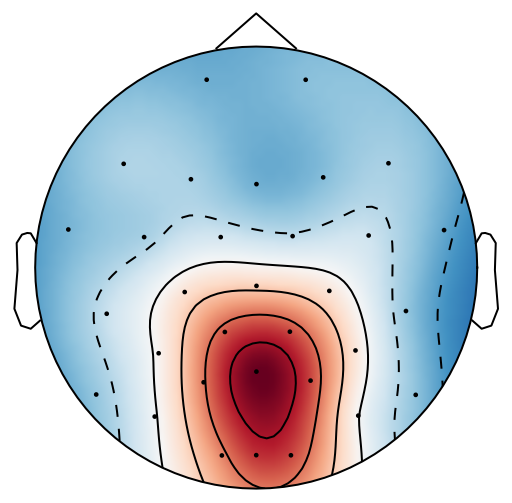}
     \includegraphics[width=.09\linewidth]{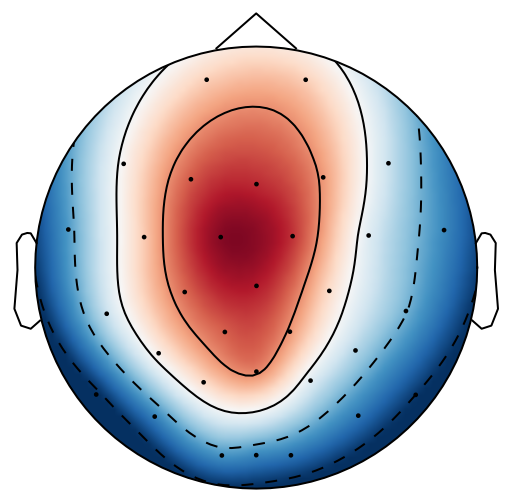}
     \includegraphics[width=.09\linewidth]{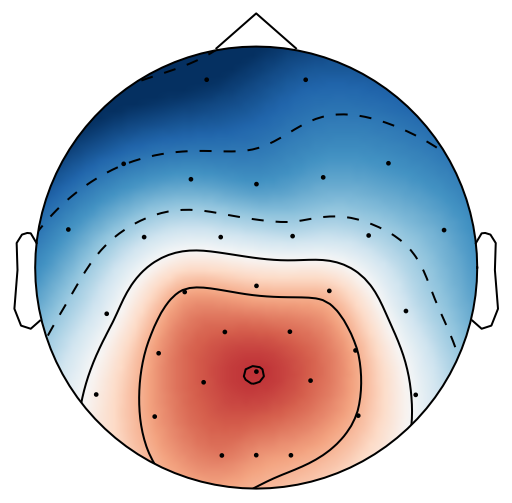}
     \includegraphics[width=.105\linewidth]{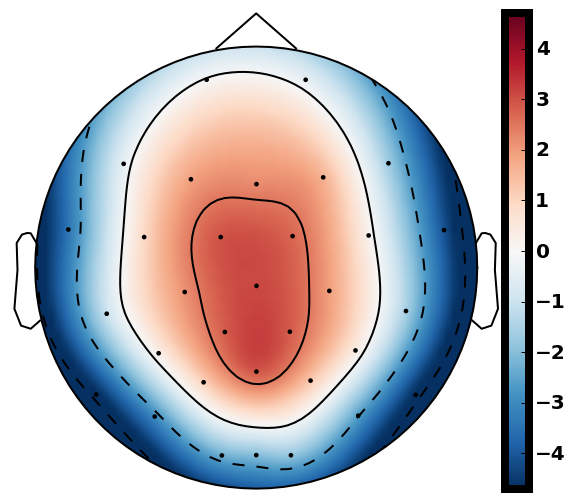}
     \label{fig:later-ERP}
	}
    \caption{ERPs for each participant corresponding to time regions of (a) 200-340 $ms$ and (b) 370-700 $ms$. These are selected for presentation to emphasize the presence of discriminative ERP-related activity in these time regions across participants, namely time regions coinciding with P200, N200 and P300 ERP activity.}
    \label{fig:two-window-spatial-pattern}
\end{figure}

The P300 is not the only ERP that is commonly encountered when using an RSVP target search paradigm. Earlier ERPs (notably the N200) are often present alongside the P3 \cite{odonnell1988n2} and can be useful in providing discriminative information for classification. Fig.~\ref{fig:two-window-spatial-pattern} shows the discriminant ERPs for 9 participants in two different time regions and it can be seen that both early time regions and later time regions generate discriminative ERP-related activity across participants. It can be noted also that the specific latencies and topographies vary across participants, hence $N_{f}$ may vary across participants for capturing target-related ERP phenomena in the RSVP paradigm.

\begin{figure}[!htp]
 \centering
	\subfloat[Spatial patterns estimated by xDAWN]{
    \centering
	 \includegraphics[width=.9\linewidth]{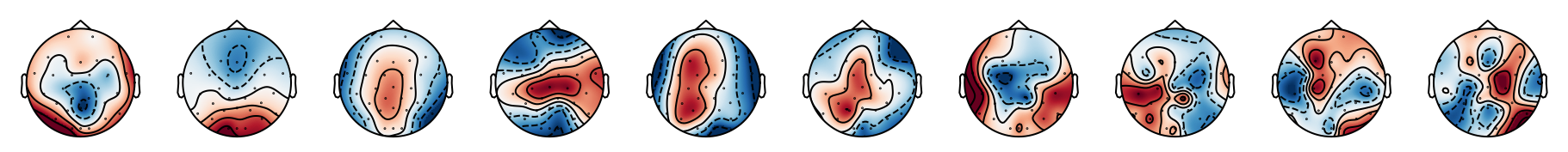}
	 \includegraphics[scale=0.2]{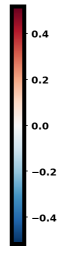}
     \label{fig:xdawn-pattern}
	}
	\hfill
    \subfloat[Spatial filters estimated by xDAWN]{
    \centering
	 \includegraphics[width=.9\linewidth]{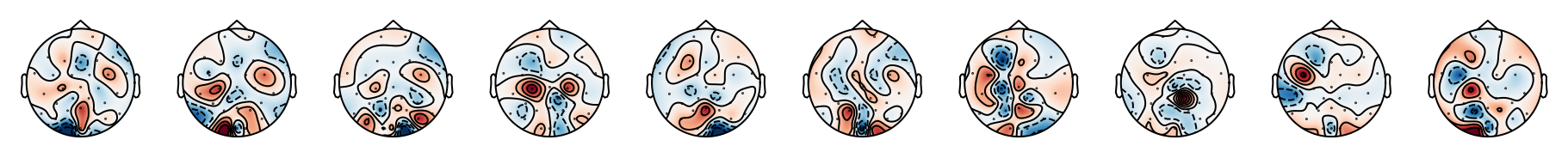}
	 \includegraphics[scale=0.2]{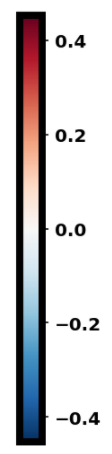}
     \label{fig:xdawn-filter}
	}
	\hfill
    \subfloat[Spatial patterns estimated by MTWLB]{
    \centering
	 \includegraphics[width=.9\linewidth]{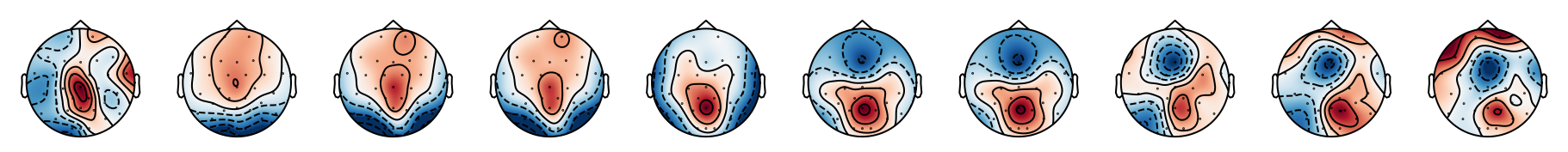}
	 \includegraphics[scale=0.2]{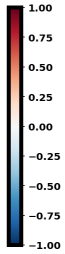}
     \label{fig:beam-pattern}
	}
    \hfill
    \subfloat[Spatial filters estimated by MTWLB]{
    \centering
	 \includegraphics[width=.9\linewidth]{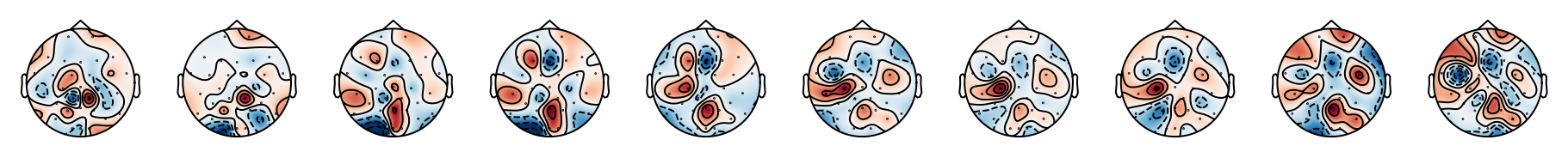}
	 \includegraphics[scale=0.2]{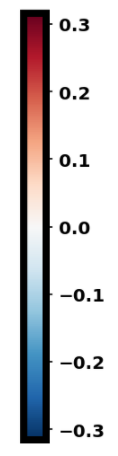}
     \label{fig:beam-filter}
	}
	\hfill
	\subfloat[Spatial patterns estimated by CSP]{
    \centering
	 \includegraphics[width=.9\linewidth]{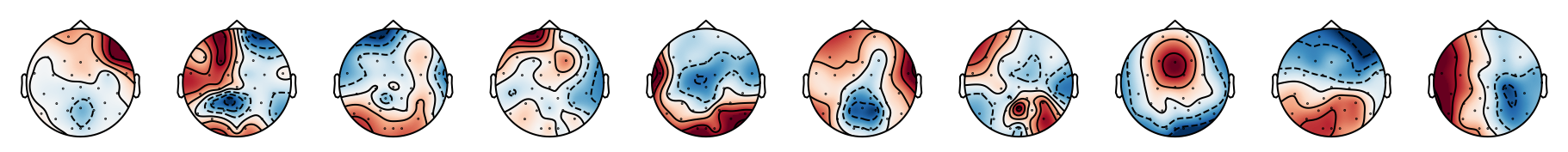}
	 \includegraphics[scale=0.2]{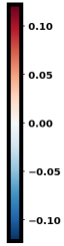}
     \label{fig:csp-pattern}
	}
    \hfill
    \subfloat[Spatial filters estimated by CSP]{
    \centering
	 \includegraphics[width=.9\linewidth]{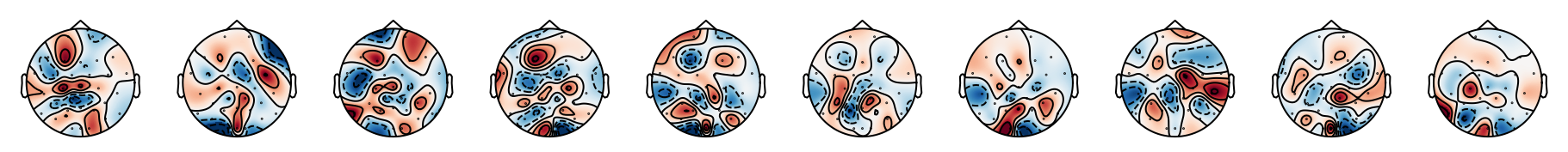}
	 \includegraphics[scale=0.2]{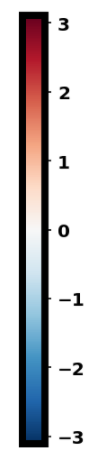}
     \label{fig:csp-filter}
	}
	
    \caption{Example of estimated spatial patterns with three spatial filtering approaches by setting $N_{f}=10$ based on 0-1 $s$ EEG epochs: \textit{Participant 1}.}
    \label{fig:SF-pattern}
\end{figure}

% changed for reviewer 1 major concern 3
From previous work \cite{cecotti2014single}, $N_{f}$ has been set to 4 for both xDAWN and CSP methods. It is difficult to determine the optimal $N_{f}$, thus we leave it as a searching hyperparameter in our pipeline. Even selection of the number of spatial filters has been stated in the area of the motor imagery BCI \cite{yang2012automatic}, we have more hyperparameters in this work. In this case, we search for the optimal number of spatial filters with other parameters together in each model, which has been specified in the Evaluation Section. It is worth reiterating again that this has not been explicitly reported upon previously in the area of RSVP. Fig.~\ref{fig:SF-pattern} shows an example of 10 spatial patterns and filters estimated with three spatial filtering approaches. It can be seen that spatial patterns estimated with the same approach are different from each other which indicates target-related ERPs span broadly over both time and space in an RSVP paradigm. Hence, a search for an optimal value of $N_{f}$ is required for best performance

\subsection{Performance Evaluation}
This work evaluated 9 different pipelines, composed of three classifiers with three spatial filtering methods. We used whole original EEG epochs for training three classifiers as measuring metrics for comparison. The AUC for each pipeline across nine participants are presented in Table~\ref{table:pipeline-result}.

\subsubsection{Performance of spatial filtering}
For the performance of three spatial filtering methods, all three classifiers with CSP pre-processing generate lower AUC score compared to those which do not use spatial filtering. This indicates that
CSP is a wholly unsuitable spatial filtering approach for RSVP-BCI. Unlike CSP, all three classifiers with MTWLB and xDAWN spatial filtering pre-processing perform better than those without spatial filtering which show the efficacy of MTWLB and xDAWN pre-processing. This result demonstrates that it is critical to select carefully the precise spatial filtering method in RSVP-BCI and an `improper' spatial filtering method may have deleterious effects and degenerate performance to the level of not using any spatial filtering. 

CSP aims to maximize the EEG variance difference between two classes. However, the single-trial ERPs variance difference is very small in RSVP paradigm between two classes and the challenge for single-trial ERPs detection is its low SNR. Here we define the `proper' spatial filtering approach for RSVP-based EEG as those methods which improve the SNR for the EEG signals. Both MTWLB and xDAWN optimize for maximized SNR and as a result both perform better than the inappropriately applied CSP method. A proper spatial filtering method can not only improve the quality of the EEG data but also reduce the computational complexity since spatial filtering can reduce the EEG dimensionality. 

\subsubsection{Performance of classifier}     
As mentioned before, CSP is not able to extract discriminant features for ERPs generated in an RSVP paradigm. Therefore, features generated by CSP have negative effects on all three classifiers compared to the EEG data without spatial filtering. From results generated by LR across  MTWLB and xDAWN and without spatial fitering, it can be seen that the performance of LR improved significantly when using non-CSP spatial filtering methods (i.e. 92.2\% for MTWLB and 92.7\% for xDAWN versus 87.5\% and 88.5\% without spatial filtering). This indicates that the quality of features has large impact on LR. For another two classifiers, spatial filtering improves the performance of LDA and BLR slightly. This indicates that LDA and BLR are more robust to the quality of features compared to LR. However, LR shows good performance if good features can be extracted by pre-processing (i.e. highest AUC score for LR with xDAWN).

\section{Discussion}
In this paper, we addressed two main issues: 1) the impact of choice of spatial filtering method on the performance of an RSVP-based BCI single trial classification task and 2) the sensitivity of different classifier's performance to the feature types produced in a typical RSVP-BCI pipeline.

Regarding the first issue, we have shown that the performance of our novel MTWLB method and the popular xDAWN method both improve classifier performance. However, the performance generated by pipelines involving CSP is worse than those without using spatial fitering. This indicates that the choice of spatial filtering method is critical for single-trial detection of ERPs in an RSVP paradigm. In the literature, we find some work which uses CSP for RSVP-based EEG \cite{yu2011common, yu2012spatio} which illustrates that CSP is considered by at least some researchers as being a suitable method for RSVP-BCI. Our results however reveal that this is not an optimal choice. By comparing the optimization criterion of each spatial filtering approach, it should be noted that MTWLB and xDAWN both use ERPs for calculating the spatial filters (i.e. ERPs difference $\mathbf{p}$ and estimated ERP response $\mathbf{D\hat{A}}$ are used for calculating $\mathbf{w}$ equation \eqref{eq:LDA-beamformer-solution} and \eqref{eq:xDAWN-solution}).

The main difference between MTWLB and xDAWN is selecting the number of spatial filters. In MTWLB, number of spatial filters is selected corresponding to the divided individual time window that means each spatial filter maximizes SNR for ERPs in the selected time window. xDAWN uses generalized eigenvalue decomposition for whole EEG epoch and eigenvectors that correspond to high eigenvalues will be selected. Therefore, the number of spatial filters $N_{f}$ refers to the number of time windows in MTWLB while it refers to the number of eigenvectors corresponding to the highest eigenvalues in xDAWN. From the classification result, the proposed MTWLB gives  similar performance compared to xDAWN. However, this proposed approach can be well-suited for generating spatial filters for those tasks that elicit ERPs in specific time regions. CSP also uses generalized eigenvalue decomposition for optimizing the spatial filter but it uses a single-trial EEG epoch instead of ERPs to calculate the covariance matrix in equation \eqref{eq:CSP-solution}. Because RSVP-based EEG has low SNR, this optimization formulation can be effected significantly by low SNR in this case. CSP origins from the motor imagery BCI in which sensor motor rhythm (SMR), a periodic EEG, is elicited in the experiment \cite{hari1997human}. The optimization criterion for CSP is maximizing the difference of variance between two classes which coincides with the property of SMR. In an RSVP paradigm, ERPs are elicited alongside the steady-state visual evoked potentials (SSVEP) and there is a very small difference of the variance between target and standard classes. The challenge for single-tiral ERP detection is its low SNR. MTWLB and xDAWN aim to improve the SNR and SSNR respectively for the reconstructed signal which overcomes the low SNR problem. Therefore, MTWLB and xDAWN are very suitable for single-trial ERP detection in an RSVP paradigm. We explored further the performance difference between MTWLB and xDAWN. We used one-way ANOVA on the mean value of three classification methods applied with MTWLB (92.2\%) and xDAWN (92.4\%) i.e. $\left[F(1, 16) = 0.004, p = 0.95\right]$, which indicates the insignificant difference of the performance between these two spatial filtering methods. This suggests that the performance of these two methods are similar for this dataset. While MTWLB and xDAWN perform similarly, MTWLB still has some advantages. First, the proposed method MTWLB gives an intuition of producing the spatial filter corresponding to the time-line. From the generated spatial filter $\mathbf{w}$ or the projected subspace $\mathbf{\Psi}$, we can see the spatial filter or projected subspace change over time. For example, the spatial filter and spatial pattern change with time in MTWLB from left to right in Fig. \ref{fig:SF-pattern}. It provides a more physiological representation of the spatial pattern and the spatial filter changing with time which the conventional spatial filtering approach is not able to represent. Second, we searched for the appropriate time window for MTWLB due to the inherent variability in ERP latencies between participants in our case. However, MTWLB can be effective for those cases in which the time region for the ERPs are known in advance. Hence, there is no need to search for the time window and the computational complexity is reduced significantly. Third, performance of xDAWN can be affected by the selected epoch length because the optimization of xDAWN is based on the whole epoch. On the contrary, MTWLB estimates the spatial pattern and covariance in the specific time window instead of the whole EEG epoch. So changing epoch length will have no effect on the performance of the MTWLB algorithm.

Regarding the second issue of the effect of features on classifier performance we have shown the performance of three classifiers in different pipelines. It can be noted that LDA and BLR outperform LR in the CSP pipeline and the pipeline without spatial filtering. This indicates that LR is more sensitive to the quality of features compared to LDA and BLR. LR is used for modeling the relationship between independent and categorical dependent variables and variable colinearities may have negative effects on estimation \cite{park2013introduction}. From the pipeline with proper spatial filtering, three classifiers perform closely to each other with  MTWLB and LR outperforming  the other two methods in the pipeline when using xDAWN. This indicates that the LR classifier performs well with informative features as input. LDA and BLR have been used more widely compared to LR in the literature since LDA performs well even without feature extraction and is simpler to implement \cite{cecotti2014single, cecotti2017best, bigdely2008brain, meng2012characterization}. Here we have shown that LR is able to generate very good performance when informative features are extracted from RSVP-based EEG.

Result in this paper partly supports the result in \cite{cecotti2014single} that spatial filtering can improve the overall performance. In this work, we performed a more comprehensive comparison of the spatial filtering pipeline. First, we included a random search \cite{bergstra2012random} for the set of hyperparameters listed in the Table \ref{table:hyperparameter-summary} in order to attain optimal performance. In the experiment, we found that the number of spatial filters can have critical impact on the final classification performance and it varies with different participants. Instead of leaving it as a specific number \cite{cecotti2014single}, we suggest to search it as a hyperparameter in the pipeline in order to optimize the performance of spatial filtering. Second, the type of spatial filtering is critical to the classification performance for different types of EEG. For example, CSP has been widely used for motor imagery based BCI \cite{ramoser2000optimal, wang2006common, lu2010regularized}, where oscillatory EEG activity is elicited in the experiment \cite{pfurtscheller2000current, pfurtscheller2001motor}. However, the classification performance of pipeline with applying CSP spatial filtering is even worse than pipelines without spatial filtering, which supports the results in \cite{cecotti2014single}. These results suggest that improper use of spatial filtering in RSVP-based BCI system can have negative impact on the classification performance and this conclusion can be extended to other types of EEG activity. On the contrary, applying the appropriate spatial filtering technique (i.e. MTWLB, xDAWN for RSVP-based EEG in this work) results in reduced computational complexity and improvement in classification performance. This work demonstrates that it is critical to choose the appropriate type of
spatial filtering for the signal processing pipeline in RSVP-based BCI systems.

\section{Conclusion}
In this work, we present a novel spatial filtering approach (MTWLB) for RSVP-based EEG. Our results demonstrate comparable performance with the leading method of xDAWN although the approach is significantly different.
Consequently the method presents a different set of optimization parameters which may make it suitable for particular RSVP-BCI implementations. Even there is no statistical significance between our proposed method and xDAWN, MTWLB presents  useful properties that lend themselves to certain RSVP-BCI performance optimizations not available via xDAWN. First, the method is more robust to EEG epoch length compared to the conventional spatial filtering approaches (e.g. xDAWN and CSP) because its optimization relies on the time window instead of whole epochs. Second, MTWLB can be more effective when knowing ERPs time region in advance because there is no need to search for the time window and the computational complexity is reduced significantly. Furthermore this work includes a thorough evaluation of single-trial classification pipelines with a number of spatial filters and classifiers in a comprehensive way using a publicly available dataset. We have shown that the selection of spatial filtering method should correspond to the nature of ERPs elicited in the task paradigm and that naive application of the approach may not produce good performance. Finally we demonstrate that even though LDA and BLR are the most prevalent classification approaches used in RSVP-based BCI research, the LR method can be even more effective for single-trial ERPs detection when good quality features are made available through the spatial filtering methods. In summary this paper should help inform designers of RSVP-BCI of an appropriate spatial filtering / classifie'r choice at design time based on results with a publicly available dataset which allows comparative benchmarking of performance.

\section{Acknowledgement}
This work is funded as part of the Insight Centre for Data Analytics which is supported by Science Foundation Ireland under Grant Number SFI/12/RC/2289.

\clearpage
\section{Appendix}

\begin{table}[!htp]
	\centering
	\caption{Hyperparameter summary for each pipeline discussed in this paper. \label{table:hyperparameter-summary}}
	\begin{tabular}{|c|c|}
		\hline
		{Pipeline}&{Hyperparameter}\\
        \hline
        $\text{MTWLB}_\text{LDA}$ & $N_{f}$ \\
		\hline
        $\text{MTWLB}_\text{BLR}$ & $N_{f}, \beta, \alpha$ \\
		\hline
        $\text{MTWLB}_\text{LR}$ & $N_{f}, \lambda$ \\
		\hline
        $\text{xDAWN}_\text{LDA}$ & $N_{f}$ \\
		\hline
        $\text{xDAWN}_\text{BLR}$ & $N_{f}, \beta, \alpha$ \\
		\hline
        $\text{xDAWN}_\text{LR}$ & $N_{f}, \lambda$\\
        \hline
        $\text{CSP}_\text{LDA}$ & $N_{f}$ \\
		\hline
        $\text{CSP}_\text{BLR}$ & $N_{f}, \beta, \alpha$ \\
		\hline
        $\text{CSP}_\text{LR}$ & $N_{f}, \lambda$\\
		\hline
        $\text{ALL}_\text{LDA}$ & $None$ \\
		\hline
        $\text{ALL}_\text{BLR}$ & $ \beta, \alpha$ \\
		\hline
        $\text{ALL}_\text{LR}$ & $\lambda$\\
		\hline
	\end{tabular}
\end{table}

\begin{table}[!htp]
	\centering
	\caption{AUC score (\%) for different pipelines across nine participants in testing session \label{table:pipeline-result}}
	\begin{tabular}{|c|c|c|c|c|c|c|c|c|c|c|}
		\hline
		\multirow{2}{*}{Pipeline}&\multicolumn{9}{c|}{Participant}& \multirow{2}{*}{Mean}\\
		\cline{2-10}
		& 1 & 2 & 3 & 4 & 5 & 6 & 7 & 8 & 9 &\\
		\hline
		$\text{MTWLB}_\text{LDA}$ &88.0  &93.8  &92.5  &96.8  &91.4  &93.7  &93.5  &90.1  &89.9  &92.2\\

		$\text{MTWLB}_\text{BLR}$ &88.0  &93.8  &92.5  &96.8  &92.2  &93.7  &93.2  &90.8  &93.3  &92.3\\
		
		$\text{MTWLB}_\text{LR}$ &88.5  &93.0  &89.8  &97.4  &91.7  &94.1  &93.3  &91.3  &90.9  &92.2\\
		
		$\text{Mean}$            &88.2  &93.5  &91.6  &97.0  &91.8  &93.8  &93.3  &90.7  &91.4  &92.2\\
		\hline
		$\text{xDAWN}_\text{LDA}$ &88.0  &93.4  &92.7  &97.3  &91.6  &94.3  &92.9  &90.6  &90.1  &92.3\\
		
		$\text{xDAWN}_\text{BLR}$ &88.5  &93.4  &92.8  &96.6  &91.6  &94.3  &92.8  &90.6  &90.1  &92.3\\
		
		$\text{xDAWN}_\text{LR}$ &87.4  &94.1  &92.9  &97.2  &91.9  &95.3  &93.8  &91.3  &90.7  &92.7\\
		
		$\text{Mean}$            &88.0  &93.6  &92.8  &97.0  &91.7  &94.6  &93.2  &90.8  &90.3  &92.4\\
		\hline
		$\text{CSP}_\text{LDA}$ &84.3  &92.8  &88.8  &96.0  &87.7  &90.9  &89.8  &76.9  &89.7  &88.5\\
		
		$\text{CSP}_\text{BLR}$ &84.9  &92.7  &88.3  &96.0  &87.9  &90.9  &89.6  &77.5  &89.7  &88.6\\
		
		$\text{CSP}_\text{LR}$ &83.8  &92.7  &85.7  &93.1  &86.6  &90.3  &89.5  &76.8  &89.0  &87.5\\
		
		$\text{Mean}$          &94.3  &92.7  &87.6  &95.0  &87.4  &90.7  &89.6  &77.1  &89.5  &88.2\\
		\hline
		$\text{ALL}_\text{LDA}$ &86.9  &93.4  &90.8  &96.7  &91.9  &94.0  &95.0  &89.4  &90.3  &92.0\\
		
		$\text{ALL}_\text{BLR}$ &88.0  &93.1  &91.4  &96.0  &91.6  &93.4  &93.8  &89.4  &90.7  &91.9\\
		
		$\text{ALL}_\text{LR}$ &84.4  &92.4  &86.0  &92.3  &87.6  &92.1  &91.5  &87.3  &82.8  &88.5\\
		
		$\text{Mean}$          &86.4  &93.0  &89.4  &95.0  &90.4  &93.2  &93.4  &88.7  &87.9  &90.8\\
		\hline	
	\end{tabular}
\end{table}

\clearpage
\bibliographystyle{tfnlm}
\bibliography{interactnlmsample}

\end{document}